\documentclass[twocolumn,prl,preprintnumbers,superscriptaddress,amsmath,amssymb,english]{revtex4-1}
\usepackage{amsmath,amssymb,amsfonts,mathrsfs}
\usepackage{amsthm}
\usepackage{CJK}
\usepackage{bm}
\usepackage{lipsum}
\usepackage{babel}
\usepackage{graphicx}
\allowdisplaybreaks
\usepackage{braket}
\usepackage[breaklinks]{hyperref}
\usepackage{color}
\hypersetup{colorlinks=true, linkcolor=blue, citecolor=blue, filecolor=blue, urlcolor=blue}
\begin{document}

\title{Photoinduced High-Chern-Number Quantum Anomalous Hall Effect from Higher-Order Topological Insulators}
\author{Xiaolin Wan}
\affiliation{Institute for Structure and Function $\&$ Department of Physics $\&$ Chongqing Key Laboratory for Strongly Coupled Physics, Chongqing University, Chongqing 400044, P. R. China}

\author{Zhen Ning}
\affiliation{Institute for Structure and Function $\&$ Department of Physics $\&$ Chongqing Key Laboratory for Strongly Coupled Physics, Chongqing University, Chongqing 400044, P. R. China}

\author{Dong-Hui Xu}
\affiliation{Institute for Structure and Function $\&$ Department of Physics $\&$ Chongqing Key Laboratory for Strongly Coupled Physics, Chongqing University, Chongqing 400044, P. R. China}
\affiliation{Center of Quantum Materials and Devices, Chongqing University, Chongqing 400044, P. R. China}

\author{Baobing Zheng}
\email{scu$_$zheng@163.com}
\affiliation{College of Physics and Optoelectronic Technology, Baoji University of Arts and Sciences, Baoji 721016, P. R. China}
\affiliation{Institute for Structure and Function $\&$ Department of Physics $\&$ Chongqing Key Laboratory for Strongly Coupled Physics, Chongqing University, Chongqing 400044, P. R. China}

\author{Rui Wang}
\email[]{rcwang@cqu.edu.cn}
\affiliation{Institute for Structure and Function $\&$ Department of Physics $\&$ Chongqing Key Laboratory for Strongly Coupled Physics, Chongqing University, Chongqing 400044, P. R. China}
\affiliation{Center of Quantum Materials and Devices, Chongqing University, Chongqing 400044, P. R. China}

\begin{abstract}
Quantum anomalous Hall (QAH) insulators with high Chern number host multiple dissipationless chiral edge channels, which are of fundamental interest and promising for applications in spintronics and quantum computing. However, only a limited number of high-Chern-number QAH insulators have been reported to date.
Here, we propose a dynamic approach for achieving high-Chern-number QAH phases in periodically driven two-dimensional higher-order topological insulators (HOTIs).
In particular, we consider two representative kinds of HOTIs which are characterized by a quantized quadruple moment and the second Stiefel-Whitney number, respectively. Using the Floquet formalism for periodically driven systems, we demonstrate that QAH insulators with tunable Chern number up to four can be achieved. Moreover, we show by first-principles calculations that the monolayer graphdiyne, a realistic HOTI, is an ideal material candidate.
Our work not only establishes a strategy for designing high-Chern-number QAH insulators in periodically driven HOTIs, but also provides a powerful approach to investigate exotic topological states in nonequilibrium cases.
%high-Chern-number
%quantum anomalous Hall (QAH) insulators have attracted intensive interest because they host multiple topologically protected chiral edge channels distinguished from the conventional QAH states. Until now, the high-Chern-number QAH insulators are restricted to magnetically doped or intrinsic magnetic topological insulator (TI) film. Here, based on the generalized BBH model, we propose a new approach to realize the high-Chern-number QAH states from 2D higher-order TIs (HOTIs) by introducing light irradiation. Employing the first-principles calculations and $k\cdot p$ effective Hamiltonian, we exhibit that the 2D HOTI graphdiyne, as a realistic material, undergoes sequential topological phase transitions from conventional QAH states to high-Chern-number QAH states under irradiation of light. By manipulating the light intensity and photon energy, the Chern number of the QAH states can be continuously changed up to 4. Furthermore, we provide a phase diagram of Chern number as a function of light intensity and photon energy to benefit the potential experimental studies. Our work not only offers a different route to realize the high-Chern-number QAH states, but also provides a realistic material to study the novel topological states.

\end{abstract}


%Use showkeys class option if keyword %display desired

\maketitle
\emph{{\color{magenta}Introduction.}}---The quantum anomalous Hall (QAH) effect, characterized by quantized Hall resistance in the absence of external magnetic field, exhibits nontrivial chiral edge states that conduct dissipationless charge current. This intriguing property facilitates its potential applications for low-power consumption
electronic devices, thus drawing considerable attention \cite{doi:10.1126/science.1187485,doi:10.1126/science.1234414,Chang2015,PhysRevLett.114.187201,RevModPhys.95.011002}. The Chern number $\mathcal{C}$, also known as the TKNN number %in condensed matter physics
and defined by the integration of the Berry curvature over the first Brillouin zone (BZ) \cite{PhysRevLett.49.405}, denotes the number of topologically protected chiral edge states, and the Hall resistance of a QAH insulator is quantized into $h/\mathcal{C}e^2$, where $h$ is the Planck's constant and $e$ denotes the charge of an electron.
%defined by the integration of the Berry curvature over the first Brillouin zone (BZ), is first employed to describe the quantum Hall effect \cite{PhysRevLett.49.405}, and can also characterize the QAH states and determines the quantized Hall resistance $h/\mathcal{C}e^2$.
Usually, the QAH state with $\mathcal{C}=1$ has been widely studied both in theories and experiments \cite{Deng2018,McIver2020,Zhao2020,Deng2021,Li2021,Guo2023}. The low-Chern-number QAH phase (i.e., $\mathcal{C}=1$), which possesses a single current transport channel, may encounter challenging issues in practical applications due to unavoidable contact resistance restrictions \cite{PhysRevLett.111.136801,Zhao2020,PhysRevLett.112.046801}.
%Conventional QAH states with $\mathcal{C}=1$, possessing single current transport channels, will face challenging problems when it turns to practical applications, for example, the restriction of unavoidable contact resistance.
In contrast, the high-Chern-number QAH states (i.e., $\mathcal{C}>1$) provide more topologically protected chiral edge channels, which can significantly improve the performance of QAH-based devices. % than the conventional QAH states.
More importantly, the high-Chern-number QAH insulators can enhance the effective breakdown current of edge states and thus provide an efficient solution to the contact resistance problem \cite{PhysRevLett.111.136801,Zhao2020,PhysRevLett.112.046801}.
%not only provide an efficient solution to the contact resistance problem in QAH-based devices, but also significantly enhance the effective breakdown current of edge states \cite{PhysRevLett.111.136801,Zhao2020,PhysRevLett.112.046801}. %As a result, such intriguing quantum states have attracted much attention in the past few years \cite{PhysRevB.85.045445,PhysRevB.105.085113,PhysRevB.103.014410,PhysRevB.105.155122,10.1093/nsr/nwaa089,PhysRevLett.111.136801,PhysRevLett.112.046801}.
However, to date, only a limited number of high-Chern-number QAH insulators have been reported in magnetically doped or intrinsic magnetic topological insulator (TI) films \cite{PhysRevB.85.045445,PhysRevB.105.085113,PhysRevB.103.014410,PhysRevB.105.155122,10.1093/nsr/nwaa089,PhysRevLett.111.136801,PhysRevLett.112.046801, PhysRevLett.129.036801, PhysRevB.107.155114,PhysRevB.93.184306}.

%the high-Chern-number QAH states are mainly proposed in the magnetically doped or intrinsic magnetic topological insulator (TI) film until now.

The QAH effect is generally suggested to be achieved in two-dimensional (2D) topological insulators (TIs) by breaking time-reversal ($\mathcal{T}$) symmetry \cite{PhysRevLett.95.226801,PhysRevB.76.045302,doi:10.1126/science.1133734,PhysRevLett.98.106803,Hsieh2008,RevModPhys.82.3045,RevModPhys.88.021004}, and consequently the $\mathcal{T}$-broken band topology indicates that chiral edge states are a necessary feature of the one-dimensional boundaries in two spatial dimensions, as dictated by the bulk-boundary correspondence.
%that possess nontrivial conducting states on their $(d-1)$-dimensional boundaries  in $d$  dimensions \cite{PhysRevLett.95.226801,PhysRevB.76.045302,doi:10.1126/science.1133734,PhysRevLett.98.106803,Hsieh2008,RevModPhys.82.3045,RevModPhys.88.021004}  by breaking time-reversal ($\mathcal{T}$) symmetry, which can remove the QAH states of one spin in 2D $\mathcal{T}$-symmetry protected TIs and leave the QAH states of another spin, consequently leading to the QAH effect.
Recently, the guiding principle for TIs has been extended to higher-order topological insulators (HOTIs), which manifest that the dimensionality of topologically gapless boundary states is lower by more than one dimension compared to that of the bulk \cite{PhysRevB.96.245115,doi:10.1126/science.aah6442,doi:10.1126/sciadv.aat0346,PhysRevLett.119.246401,PhysRevLett.119.246402}.
%Specifically, the $n$-th order TIs in $d$  dimensions have gapless states on the ($d-n$)-dimensional boundaries ($n\geq2$ corresponds to HOTIs).
The novel bulk-boundary correspondence has propelled  HOTIs into the forefront of research in the field of topological phases of matter \cite{PhysRevLett.119.246401,PhysRevLett.119.246402,PhysRevB.97.205136,PhysRevB.97.205135,PhysRevLett.124.036803, Schindler2018,Noguchi2021,doi:10.1126/sciadv.aat0346,PhysRevLett.122.256402,PhysRevLett.123.256402,Tang2019,Tang2019NP,PhysRevLett.123.186401, PhysRevB.107.035128,PhysRevB.104.205117}.
%that the topologically protected surface (edge) states associate with the topology of bulk band structure, i.e., so-called bulk-boundary correspondence, has been generalized to higher order in crystalline insulators using the theory of dipole moments \cite{PhysRevB.96.245115,doi:10.1126/science.aah6442}, and consequently the higher-order topological insulators (HOTIs) were proposed \cite{doi:10.1126/sciadv.aat0346,PhysRevLett.119.246401,PhysRevLett.119.246402}. HOTIs characterize with insulating bulk and gapped surfaces/edges, but possess topologically protected conducting states at 0D corners or 1D hinges. Specifically, the $n$-th order TIs in $d$  dimensions have gapless states on the ($d-n$)-dimensional boundaries ($n\geq2$ corresponds to HOTIs).  Most importantly, besides the study based on the model Hamiltonian and symmetry arguments \cite{PhysRevLett.119.246401,PhysRevLett.119.246402,doi:10.1126/sciadv.aat0346,PhysRevB.97.205136,PhysRevB.97.205135,PhysRevLett.124.036803}, the realistic materials of HOTIs have been confirmed experimentally \cite{Schindler2018,Noguchi2021}, and several theoretical proposals are extensively suggested to complement experiment \cite{doi:10.1126/sciadv.aat0346,PhysRevLett.122.256402,PhysRevLett.123.256402,Tang2019,Tang2019NP,PhysRevLett.123.186401}. This further promotes the advancement of HOTIs.
%Following this essential expansion of topological states of matter, great attention has been paid to these exotic quantum states, it is natural to ask whether the high-Chern-number QAH states can be achieved in HOTIs.
The $\mathcal{T}$-symmetry plays a vital role in protecting the higher-order boundary states in HOTIs associated by the crystalline symmetries \cite{doi:10.1126/sciadv.aat0346,PhysRevLett.119.246401,PhysRevLett.119.246402}, even though crystalline symmetries are not essential \cite{PhysRevLett.119.246401,PhysRevX.9.011012, PhysRevLett.125.166801}. Therefore, in light of the fundamental expansion of topological states of matter, it is imperative to investigate topological phase transitions by introducing $\mathcal{T}$-symmetry breaking in HOTIs. To be specific, there are currently at least two unresolved issues about this problem: (I) the connection between the QAH and high-Chern-number QAH states and the $\mathcal{T}$-broken HOTIs; (II) the realization approach and material candidate, if Issue (I) is established.

Motivated by the above issues, in this work, we propose a strategy for achieving high-Chern-number QAH insulators from 2D $\mathcal{T}$-invariant HOTIs irradiated by circularly polarized light (CPL). We focus on two representative types of 2D HOTIs whose bulk topology can be respectively measured by a quantized quadruple moment~\cite{PhysRevB.96.245115,doi:10.1126/science.aah6442} and the second Stiefel-Whitney number~\cite{2rdSWnumber}. The latter one has been realized in 2D graphdiyne~\cite{PhysRevLett.123.256402,lee2020two}. It is noteworthy that, besides the application of external magnetic fields and magnetic dopant atoms, the utilization of periodically driven light fields is considered to be an effective approach for breaking $\mathcal{T}$-symmetry and exploring exotic topological phases out of equilibrium \cite{Lindner2011,doi:10.1126/science.1239834,McIver2020,He2019,Ito2023,Zhou2023}. Within the framework of the Floquet formalism, we reveal that the high-Chern-number QAH phase is derived from the light-induced multiple band inversion collaborated by higher-order band topology. The Chern number is continuously tunable from 1 to 4 by changing the light intensity and photon energy.

\emph{{\color{magenta}High-Chern-number QAH state from the quadruple insulator.}}---We start with the Benalcazar-Bernevig-Hughes (BBH) Hamiltonian for a quantized quadrupole insulator described as
\cite{PhysRevB.96.245115,doi:10.1126/science.aah6442}
\begin{equation}\label{Eq.1}
\begin{split}
H(\mathbf{k})=&\lambda\sin(k_y)\Gamma_{1}+[\gamma_y+\lambda\cos(k_y)]\Gamma_{2} \\
                 &+\lambda\sin(k_x)\Gamma_{3}+[\gamma_x+\lambda\cos(k_x)]\Gamma_{4},
\end{split}
\end{equation}
in which $\mathbf{k}=(k_x,k_y)$ is the wave vector, $\lambda$ and $\gamma$ represent two kinds of hopping strengths, $\Gamma_i=-\tau_{2}\sigma_{i}$ ($i=1, 2$, and $3$), and $\Gamma_4=\tau_{1}\sigma_{0}$ (Here, $\tau$ and $\sigma$ denote two sets of Pauli matrices for different degrees of freedom). For simplicity and without loss of generality, we hereafter assume $\lambda=1$ and $\gamma_x=\gamma_y=\gamma$. We here mainly focus on the topological properties at the $\Gamma=(0,0)$, and expand Eq. (\ref{Eq.1}) to the $k$-quadratic order around $\Gamma$ to obtain the low-energy effective model $H(\mathbf{k})=k_y\Gamma_{1}+M_y\Gamma_{2}+k_x\Gamma_{3}+M_x\Gamma_{4}$, where $M_{x(y)}=1+\gamma-\frac{1}{2}k_{x(y)}^2$.
To study the influence of light field on the quadrupole insulator, we apply a time-periodic and spatially homogeneous CPL that can be described by vector potential of $\mathbf{A}(t)=A[\cos(\omega t), \eta\sin(\omega t), 0]$ for polarization in the $x$-$y$ plane, in which A is the amplitude, $\omega=2\pi/T$ is frequency of light, and $\eta=\pm$ are for right- and left-handed CPL that rotates counterclockwise and clockwise in the $x$-$y$ plane, respectively. The coupling of the time-periodic CPL on the effective Hamiltonian can be described by Peierls substitution, resulting time-dependent Hamiltonian $H(\mathbf{k},t)=H(\mathbf{k}+\frac{e}{\hbar}\mathbf{A(t)})$~\cite{PhysRevA.27.72,PhysRevLett.110.200403,Hubener2017}. Then, we expand the CPL driven Hamiltonian by discrete Fourier series and obtain $H(\mathbf{k},t)=\sum_{m}H_{m}(\mathbf{k})e^{-im\omega t}$. The Fourier components $H_{m}(\mathbf{k})$ read
\begin{equation}\label{Eq.2}
H_{m}(\mathbf{k})=\frac{1}{T}\int_{0}^{T}e^{im\omega t}H(\mathbf{k}+\frac{e}{\hbar}A(t))dt.
\end{equation}
\begin{figure}
    \centering
    \includegraphics[width=\linewidth]{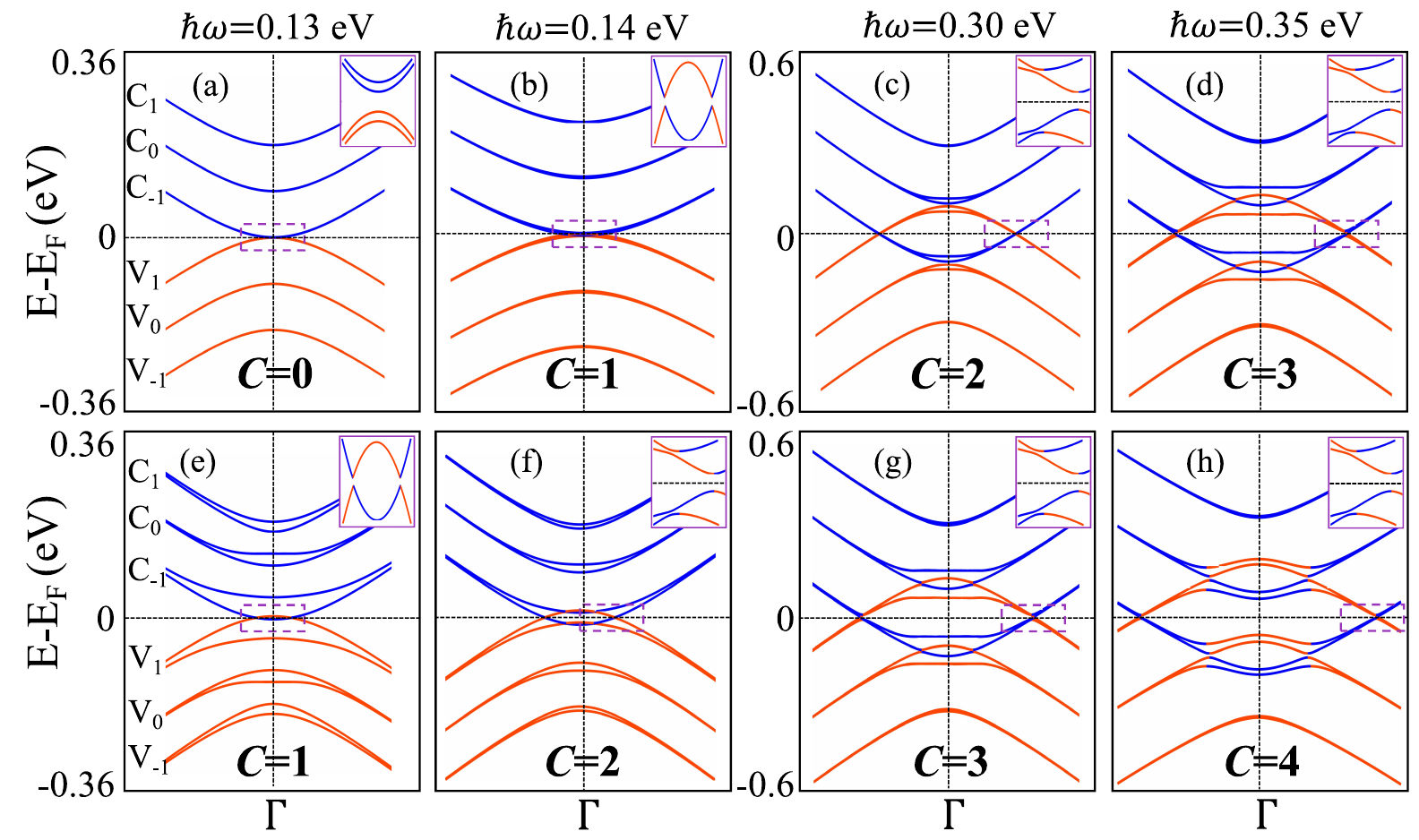}
    \caption{The evolution of energy spectrum of quadrupole insulator  under irradiation of CPL with the increasing light intensity $eA/\hbar$ and photon energy  $\hbar\omega$, in which the corresponding calculated Chern number are marked, C$_{1,0,-1}$ and V$_{1,0,-1}$ represent the photon-dressed bands near the Fermi level, respectively. In (a)-(d), the light intensity $eA/\hbar$ is set to 0.03 $\mathrm{\AA}^{-1}$, while $eA/\hbar=0.08$ $\mathrm{\AA}^{-1}$  in (e)-(h). The photon energy of CPL in the same vertical column, as labelled on the top of corresponding columns, are 0.13, 0.14, 0.30, 0.35 eV, respectively. The insets enlarge the bands in the dashed box near the Fermi level. Here, we have set $|\gamma/\lambda|=0.9$ for all the calculations to guarantee the nontrivial quadrupole phase of the BBH model.
    \label{FIG1}}
\end{figure}
Based on Eq. (\ref{Eq.2}), we can construct an effective static Hamiltonian $H(\mathbf{k},\omega)$ in the infinite extended Hilbert space of multi-photon components [see details in Supplemental Material (SM)]~\cite{Hubener2017,https://doi.org/10.48550/arxiv.2003.08252,SM}. After careful inspection of the convergence of the Floquet quasienergy spectrum, we find that the first-order truncation is accurate enough to describe the irradiation of CPL, and the high-order terms that correspond to the contributions of two-photon and beyond have negligible influence on the Floquet quasienergy spectrum. Therefore, we here truncate the infinite Floquet bands to first order in the following discussion. The considered terms read,
\begin{equation}\label{Eq.3}
\begin{split}
H_{0}(\mathbf{k})&= k_y\Gamma_{1}+(M_y-B^2)\Gamma_{2}+k_x\Gamma_{3}+(M_x-B^2)\Gamma_{4} \\
H_{\pm1}(\mathbf{k})&= B[\Gamma_{3}-k_{x}\Gamma_{4}\pm i(\Gamma_1-k_y\Gamma_2)\eta] \\
H_{\pm2}(\mathbf{k})&= \frac{B^2}{2}(\Gamma_{2}-\Gamma_{4}),
\end{split}
\end{equation}
where $B=\frac{Ae}{2\hbar}$.
By diagonalizing the truncated Floquet Hamiltonian $H(\mathbf{k},\omega)$ in the momentum and frequency space, we can obtain Floquet quasienergy bands of the quadrupole insulator around $\Gamma$ under irradiation of CPL. We here mainly present the effect of left-handed CPL on the quadrupole insulator and will no longer show the case of right-handed CPL any more because it has equivalent effect on the quadrupole insulator as that of left-handed CPL. As is well known, our employed BBH model exhibits a trivial quadrupole phase without gapless edge states for $|\gamma/\lambda|>1$, while a nontrivial quadrupole phase with quantized corner states for $|\gamma/\lambda|<1$. Therefore, we have set $|\gamma/\lambda|=0.9$ to guarantee that the CPL is incorporated into the nontrivial quadrupole phase.

The calculated Floquet quasienergy bands with increasing  light intensity $eA/\hbar$  and and photon energy  $\hbar\omega$ are shown in Fig. \ref{FIG1}. In the absence of irradiation of CPL, the energy spectrum characterizes with a direct band gap between the conduction band minimum (CBM) and the valence band maximum (VBM) at $\Gamma$, and exhibits corner zero-modes as expected \cite{doi:10.1126/science.aah6442}. When the CPL is introduced, the originally well-separated CBM and VBM respectively move downward and upward, leading to the closing and reopening of band gap, forming the inverted bands. As shown in Fig. \ref{FIG1}(a) (with light intensity $eA/\hbar=0.03 $ $\mathrm{\AA}^{-1}$ and photon energy $\hbar\omega=0.13$ eV), the Floquet quasienergy bands still keep  trivial insulating state because of the relatively small photon energy. As we increase the photon energy ($\hbar\omega=0.14$ eV), one can evidently see the band inversion emerges between the Floquet-Bloch bands near the Fermi level (i.e., C$_{-1}$ and V$_{1}$), shown in Fig. \ref{FIG1}(b). To characterize the topological property, we can calculate the Chern number $\mathcal{C}$ of the Floquet system, which can be given by an integral of the Berry curvature of the occupied bands \cite{WeijieWang66701}. The calculated Chern number $\mathcal{C}=1$ further confirms the nontrivial band topology, and demonstrates that the nontrivial quadrupole insulator evolves into a quantum anomalous Hall (QAH) insulator under irradiation of CPL.

More importantly, continuous increase in photon energy drives further evolution of the photon-dressed band structure and induces multiple band inversion for the quadrupole insulator. As shown in Figs. \ref{FIG1}(c) and (d),  the inversion of Floquet-Bloch bands appears twice and three times, characteristic of  higher Chern number $\mathcal{C}=2$ and $\mathcal{C}=3$, respectively. That is to say, under the irradiation of CPL, we can obtain the high-Chern-number QAH states from a quadrupole insulator, and the Chern number can be continuously manipulated by changing the photon energy. Meanwhile, by tuning the light intensity  at the given photon energy, the Chern number of the Floquet system can be also continuously changed. For example, at certain photon energy $\hbar\omega=0.14$ eV, the corresponding Chern number changes continuously from 1 to 2 as the light intensity increases from 0.03 $\mathrm{\AA}^{-1}$ to 0.08 $\mathrm{\AA}^{-1}$ [compared Fig. \ref{FIG1}(b) with Fig. \ref{FIG1} (f)]. Moreover, it is noteworthy  that the maximum Chern number of the Floquet system that we can obtain is 4, as shown in Fig. \ref{FIG1} (h) with $eA/\hbar=0.08$ $\mathrm{\AA}^{-1}$ and $\hbar\omega=0.35$ eV. In short, utilizing the BBH model and Floquet theory, we demonstrate that the 2D quadrupole insulator evolves into the Floquet QAH insulator with tunable high Chern number under the irradiation of CPL.

\emph{{\color{magenta}High-Chern-number QAH state from a HOTI with second Stiefel-Whitney number and its realization in graphdiyne.}}---Next, we show that the photoinduced high-Chern-number QAH states can also occur in a HOTI with second Stiefel-Whitney number. This type of HOTI was realized in 2D graphdiyne~\cite{PhysRevLett.123.256402,lee2020two}, whose bulk topology can be revealed by  a four-band $k\cdot p$ Hamiltonian as \cite{PhysRevLett.123.256402,PhysRevLett.128.026405}
%from the HOTIs
%We also employ  a four-band $k\cdot p$ Hamiltonian described as follow \cite{PhysRevLett.123.256402,PhysRevLett.128.026405}
\begin{equation}\label{Eq.11}
\begin{split}
H=&(-\Delta+mk_{x}^{2}+mk_{y}^{2})\sigma_{z}\tau_{0}+vk_{x}\sigma_{x}\tau_{x}+vk_{y}\sigma_{x}\tau_{y} \\
&+\lambda(k_{x}^{2}-k_{y}^{2})\sigma_{z}\tau_{x}+2 \lambda k_{x}k_{y}\sigma_{z}\tau_{y}.
\end{split}
\end{equation}
%to further confirm the photoinduced high-Chern-number QAH states from the HOTIs.
The calculation details and related results are included in the SM \cite{SM}.
As shown in Fig. S1 \cite{SM}, we can see that multiple band inversion of Floquet bands emerges and the 2D HOTI with the second Stiefel-Whitney number evolves into the Floquet QAH insulator with tunable Chern number under the irradiation of CPL, which is quite similar to the results obtained from the quadrupole insulator described by BBH model.

%Here, it should be noted that achieving high Chern-number QAH states is also possible in the trivial quadrupole phase (i.e., $|\gamma/\lambda|>1$) subjected to CPL. However, the practical realization of this novel state in the trivial quadrupole phase remains challenging due to the overlarge light intensity and photon energy required by our calculations. In addition, for the periodically driven systems in two dimensions, it is also possible to realize the robust chiral edge states even if the Chern numbers of all the bulk Floquet bands are zero \cite{PhysRevX.3.031005}.

\begin{figure}
    \centering
    \includegraphics[scale=0.187]{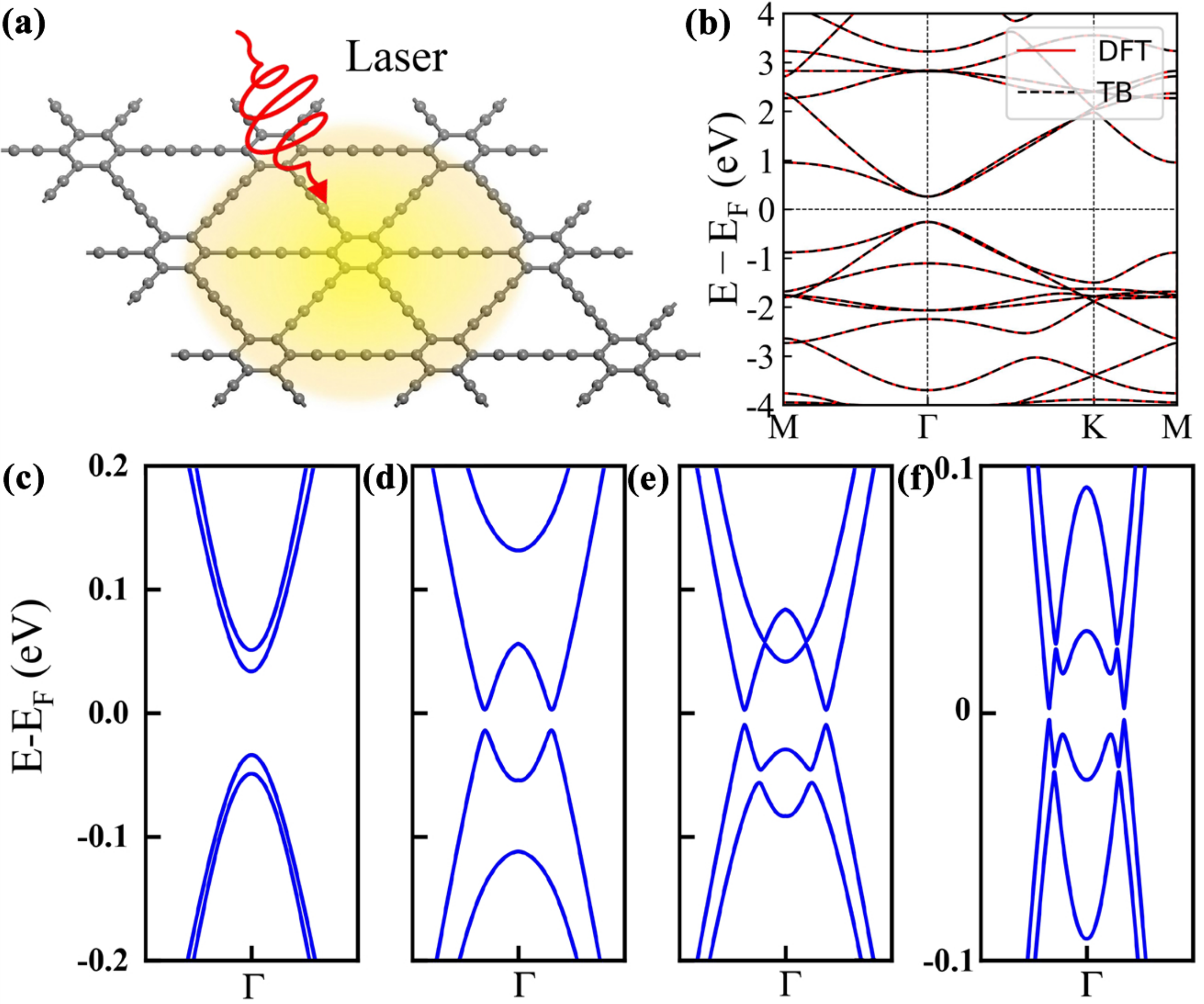}
    \caption{(a) Schematic of irradition of CPL on graphdiyne. (b) The band structures of graphdiyne obtained by DFT calculation (red solid lines) and Wannier TB Hamiltonian (black dashed lines) without irradition of CPL. (c)-(f) The Floquet band structures of graphdiyne around $\Gamma$ under irradition of CPL with different light intensity $eA/\hbar$ and photon energy $\hbar\omega$: In (c), with the parameters $eA/\hbar=0.01$ $\mathrm{\AA}^{-1}$ and $\hbar\omega=0.2$ eV, the Floquet bands are trivial; In (d), the Floquet bands with $eA/\hbar=0.04$ $\mathrm{\AA}^{-1}$ and $\hbar\omega=0.2$ eV exhibit band inversion with $\mathcal{C}=1$ near the Fermi level; In (e), the Floquet bands with $eA/\hbar=0.03$ $\mathrm{\AA}^{-1}$ and $\hbar\omega=0.31$ eV evolve into the high-Chern-number states with $\mathcal{C}=2$; With the parameters $eA/\hbar=0.02$ $\mathrm{\AA}^{-1}$ and $\hbar\omega=0.4$ eV in (f), the Chern number of the Floquet system increases to 3.
     \label{FIG2}}
\end{figure}

The material realization will greatly facilitate the potential applications of photoinduced QAH states from 2D HOTIs. Considering that 2D graphdiyne was experimentally synthesized \cite{B922733D}, by employing first-principles calculations and a tight-binding (TB) model combined with Floquet theory, we next take it as an example to confirm the photoinduced topological phase transition from the HOTI to high-Chern-number QAH states in 2D graphdiyne.
%The experimentally synthesized graphdiyne \cite{B922733D}, respecting D$_{6h}$ symmetry,  was demonstrated to possess topologically protected 0D corner states.
%Therefore, by employing the first-principles calculations and a tight-binding (TB) model associated with Floquet theory, we next take the realistic 2D HOTI material graphdiyne as an example to realize and further confirm the high-Chern-number QAH states.
Due to the negligible spin-orbital coupling of graphdiyne, we treat it as a spinless system in the following discussions.
For first-principles calculations %including structural relaxation and electronic band structure,
we adopted Perdew-Burke-Ernzerhof exchange-correlation functional within the framework of density functional theory (DFT) as implemented in the Vienna Ab \textit{initio} Simulation Package \cite{PhysRev.136.B864,PhysRevB.54.11169,PhysRevLett.77.3865}. The other calculated details can be found in the SM \cite{SM}. Figure \ref{FIG2}(a) show the optimized crystal structure of graphdiyne constructed from \emph{sp}- and \emph{sp}$^2$-hybridized carbon atoms. The calculated lattice constant is 9.46 $\mathrm{\AA}$, which shows excellent consistency with previous experimental and theoretical results \cite{doi:10.1063/1.4993916,PhysRevB.58.11009,PhysRevLett.123.256402}. The obtained band structure, as illustrated in Fig. \ref{FIG2}(b),  shows that it possesses a direct band gap of 0.47 eV at $\Gamma$, which agrees well with the previous theoretical result of 0.46 eV \cite{Long2011}.

\begin{figure}
    \centering
    \includegraphics[scale=0.082]{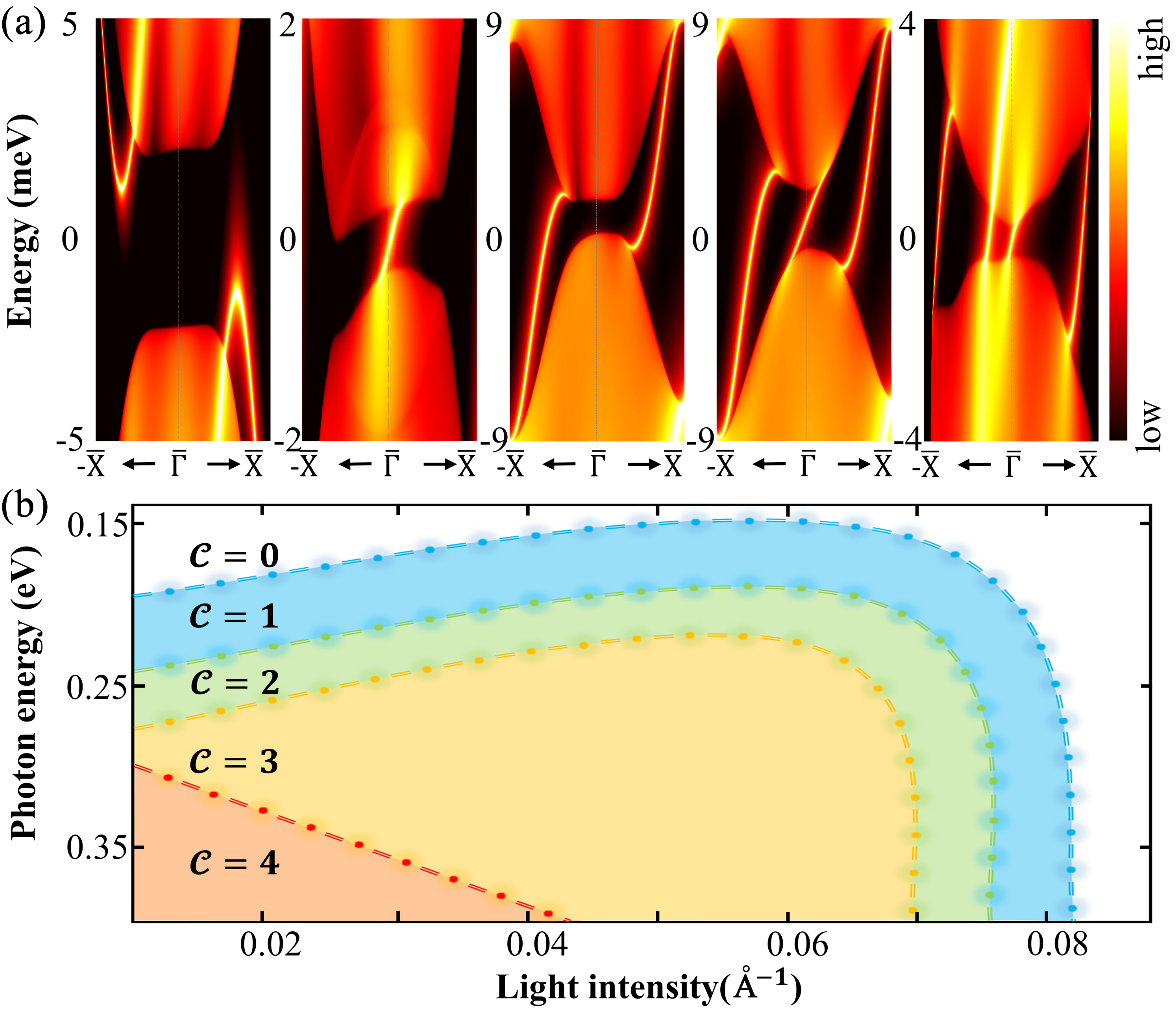}
    \caption{Edge states and phase diagram of graphdiyne under the irradiation of CPL. (a) The edge states of graphdiyne along high-symmetry zigzag edge $-\overline{\mathrm{X}}\leftarrow\overline{\Gamma}\rightarrow\overline{\mathrm{X}}$ under the irradiation of CPL with different light intensity $eA/\hbar$ and photon energy $\hbar\omega$: The leftmost panel shows the trivial states of graphdiyne corresponding to $\mathcal{C}=0$ with $eA/\hbar=0.06$ $\mathrm{\AA}^{-1}$ and $\hbar\omega=0.17$ eV; The second panel with $eA/\hbar=0.05$ $\mathrm{\AA}^{-1}$ and $\hbar\omega=0.17$ eV exhibits the edge states with single chiral edge channels; The middle panel with $eA/\hbar=0.07$ $\mathrm{\AA}^{-1}$ and $\hbar\omega=0.27$ eV corresponds to the nontrivial states of graphdiyne with $\mathcal{C}=2$; The edge states of graphdiyne with $\mathcal{C}=3$ are represented by the fourth panel ($eA/\hbar=0.069$ $\mathrm{\AA}^{-1}$ and $\hbar\omega=0.27$ eV); The rightmost panel ($eA/\hbar=0.02$ $\mathrm{\AA}^{-1}$ and $\hbar\omega=0.34$ eV) represents the edge states with four chiral edge channels for graphdiyne. (b) The Chern number phase diagram of graphdiyne under the irradiation of CPL as a function of light intensity and photon energy.
     \label{FIG3}}
\end{figure}
Based on the obtained band structure, we established a real-space Wannier TB Hamiltonian for graphdiyne. Note that, in order to ensure the accuracy and reliability of our following analysis, the band structures calculated by the TB Hamiltonian should be in good agreement with those computed by first-principle calculations, shown in Fig. \ref{FIG2}(b). Employing the Floquet theorem, we can introduce the CPL in the real-space TB Hamiltonian in the Wannier function basis and obtain an effective static Hamiltonian $H(\mathbf{k},\omega)$ in the frequency and momentum space. The further details for $H(\mathbf{k},\omega)$ can be found in the SM \cite{SM}. Here, we also consider the first-order truncation for the infinite Floquet Hamiltonian after careful inspection of the convergence of the results.

According to the truncated $H(\mathbf{k},\omega)$, we computed the band structure of 2D HOTI graphdiyne under the irradiation of CPL as implemented in WannierTools code \cite{WU2017}. The calculated band structures around $\Gamma$ under various light intensity and  photon energy of CPL are shown in Figs. \ref{FIG2}(c)-\ref{FIG2}(f). Evidently, the bands around $\Gamma$ can be inverted by tuning the light intensity and photon energy, and multiple band inversion occurs as expected. Moreover, the Chern number can be manipulated continuously from trivial state $\mathcal{C}=0$ to nontrivial QAH state with $\mathcal{C}=3$. It is worthwhile to mention the higher Chern number larger than 3 can also be realized in graphdiyne by further tuning light intensity and  photon energy.
%Our results of realistic material graphdiyne calculated by first-principles calculations are in good accordance with those obtained by BBH model, further confirming the photoinduced QAH states with tunable Chern number from 2D HOTIs.

To demonstrate the topological properties of the photoinduced QAH states, we next show the hallmark of the QAH phases with tunable Chern number by depicting their edge states. Figure \ref{FIG3}(a) illustrates the edge states of graphdiyne under the irradiation of CPL along zigzag direction $-\overline{\mathrm{X}}\leftarrow\overline{\Gamma}\rightarrow\overline{\mathrm{X}}$. One can see that the number of the chiral edge states with different light intensity and  photon energy show good agreement with the DFT band calculations, further supporting the fact that we can obtain the QAH states by irradiating CPL on a 2D HOTI. Moreover, by manipulating the light intensity and  photon energy, we can realize the high-Chern-number QAH states up to $\mathcal{C}=4$ for graphdiyne, which produces multiple topologically protected chiral edge current channels in contrast to convectional QAH insulators. It is worth noting that graphdiyne, as a 2D realistic HOTI,  is the first HOTI platform to achieve the high-Chern-number QAH states under the irradiation of CPL. This new approach is expected to counteract the drawback of contact resistance and may greatly facilitate the potential applications of QAH insulator

The phase diagram can completely summarize the parameter regimes of different phases and gain a deep insight into the topological phase transition. Therefore, based on the $H(\mathbf{k},\omega)$ obtained from first-principles calculations, we obtain the phase diagram of Chern number for graphdiyne as a function of light intensity and photon energy of CPL, as shown in Fig. \ref{FIG3}(b). We can find that there are five distinct phase regimes corresponding to the continuously changed Chern number (ranging from 0 to 4). With relatively small light intensity and photon energy, the Chern number increases from 0 to 4 with the increasing light intensity and photon energy. However, when the light intensity and photon energy is relatively high, it is found that further increase of light intensity and photon energy will suppress the high-Chern-number QAH states and lead to the decrease of Chern number rather than further increase it. This can serve as a useful guide for probing this novel topological states.

\emph{{\color{magenta}Summary.}}---In summary, we propose a dynamic approach for achieving high-Chern-number QAH phases in periodically driven 2D HOTIs irradiated by CPL. To elucidate this, we focus on two representative types of 2D HOTIs whose bulk topology can be depicted by a quantized quadruple moment and the second Stiefel-Whitney number. We respectively choose the BBH model and the 2D graphdiyne to exhibits the photoinduced topological phase transition.  This phase transition shows strong universality, i.e., the high-Chern-number QAH phase is derived from the light-induced multiple band inversion collaborated by higher-order band topology. The Chern number is continuously tunable from 1 to 4 by changing the light intensity and photon energy. Importantly, the first-priciples calculations for the realistic material graphdiyne will greatly favor the experimental verification and further potential applications. Our work not only establishes a strategy for designing high-Chern-number QAH insulators in periodically driven HOTIs, but also provides a powerful approach to investigate exotic topological states in nonequilibrium cases.

\emph{{\color{magenta}Acknowledgements.}}---This work was supported by the National Natural Science Foundation of China (NSFC, Grants No. 12222402, No. 11974062, No. 12147102, and No. 12047564), the Chongqing Natural Science Foundation (Grants No. 2023NSCQ-JQX0119), the Natural Science Basic Research Program of Shaanxi (Grant No. 2022JM-001), and the Beijing National Laboratory for Condensed Matter Physics.

%\bibliographystyle{apsrev4-2}

%\bibliography{ref}
%apsrev4-2.bst 2019-01-14 (MD) hand-edited version of apsrev4-1.bst
%Control: key (0)
%Control: author (72) initials jnrlst
%Control: editor formatted (1) identically to author
%Control: production of article title (-1) disabled
%Control: page (0) single
%Control: year (1) truncated
%Control: production of eprint (0) enabled
%

\end{document}